\documentclass[
 reprint,
superscriptaddress,
 amsmath,amssymb,
 aps,
 pra,
]{revtex4-2}

\usepackage{graphicx}
\usepackage{float}
\usepackage{dcolumn}
\usepackage{bm}
\usepackage{bbm}
\usepackage{braket}
\usepackage{xcolor}
\usepackage[T1]{fontenc}
\usepackage{enumerate,enumitem}
\usepackage{booktabs}
\usepackage{physics}
\usepackage{amsmath}
\usepackage{appendix}
\usepackage{hyperref}

\begin{document}

\title{One-clean-qubit spectroscopy of simulated Kitaev chains}
\author{Z.~M.~McIntyre}
\affiliation{Department of Physics, University of Basel, Klingelbergstrasse 82, 4056 Basel, Switzerland}
\email{zoe.mcintyre@unibas.ch}
\author{Daniel Loss}
\affiliation{Department of Physics, University of Basel, Klingelbergstrasse 82, 4056 Basel, Switzerland}
\affiliation{Physics Department, King Fahd University of Petroleum and Minerals, 31261, Dhahran, Saudi Arabia}
\affiliation{Quantum Center, KFUPM, Dhahran, Saudi Arabia}
\affiliation{RDIA Chair in Quantum Computing}
\date{\today}
\begin{abstract}
    Spin qubits in gate-defined quantum dots provide a highly programmable platform for simulating condensed-matter phenomena. In this work, we introduce a digital-analog quantum simulation protocol for extracting the single-particle spectrum of a Kitaev chain. The Kitaev chain is mapped onto qubits via the standard Jordan-Wigner transformation and implemented  as a drive-engineered, $N$-site transverse-field Ising model (TFIM) in a linear array of quantum dots. We show that periodically toggling the analog-simulation parameters conditioned on the state of a control qubit causes the dynamics of this control qubit to stroboscopically match the output of the one-clean-qubit (DQC1) model of computation, thereby yielding the full spectrum of the TFIM from measurements of a single spin. Classical postprocessing can then be used to isolate the $N$ single-particle energies of the Kitaev chain from the $2^N$ eigenenergies of the TFIM. By varying the strength of the Rabi drive used to engineer the synthetic transverse field, the spectral signature of the crossover from the trivial to the topological regime of the Kitaev chain could then be mapped out with measurements of just one spin.
\end{abstract}
\maketitle

Gate-defined quantum-dot arrays have long been recognized as a versatile platform for quantum simulation.
Experimentally, such arrays have been used for simulation of the Fermi--Hubbard model~\cite{hensgens2017quantum}, Nagaoka ferromagnetism~\cite{dehollain2020nagaoka}, antiferromagnetic Heisenberg chains~\cite{van2021quantum}, and minimal realizations of the Kitaev chain~\cite{dvir2023realization}. They have also been used to probe excitonic transport~\cite{hsiao2024exciton}, to simulate quantum walks of magnon and triplon excitations~\cite{farina2025site}, and to demonstrate many-body Ramsey interferometry for the purpose of mapping out the low-energy sector of a strongly interacting spin system~\cite{jirovec2026many}. These experimental demonstrations of using quantum dots to investigate complex physical phenomena are complemented by significant recent advances in the control of spin qubits for gate-based quantum computing~\cite{burkard2023semiconductor,mcintyre2026theory}, including the demonstration of a six-qubit quantum circuit~\cite{fernandez2026running} and simultaneous control of multiple qubits in an 18-dot   array~\cite{dijkema2026simultaneous}. 

Independent of the physical platform being used, quantum simulation requires protocols for extracting physically meaningful quantities from the system being simulated. In many cases, measuring order parameters for the purpose of distinguishing between different quantum phases may require a measurement of all qubits involved in the simulation. The ordered and disordered phases of the transverse-field Ising model, for instance, can be distinguished through the average magnetization, which can be obtained by measuring the expectation value of all spins along the direction of the exchange interaction~\cite{islam2011onset,kim2023evidence}. Site-averaged autocorrelation functions can also be used to probe many-body localization~\cite{nagao2026probing} or to verify the presence of the period-doubled oscillations characteristic of a discrete time-crystalline phase~\cite{switzer2026realization}. In the quantum-dot devices currently being fabricated for spin-based quantum computing, it is typically not possible to perform a simultaneous measurement of all spins due to the limited number of charge-sensor dots. Simulation protocols that extract useful information from measurements of a single spin may therefore find application in near-term devices.

\begin{figure}
    \centering
    \includegraphics[width=0.9\linewidth]{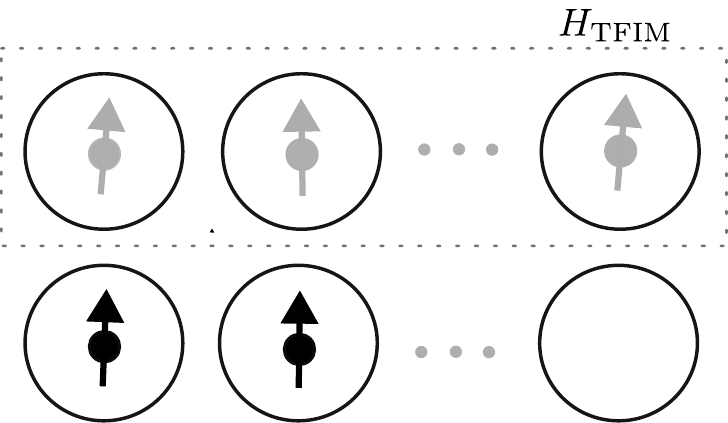}
    \caption{Schematic of a $2\times N$ array of quantum dots. The Hamiltonian $H_{\mathrm{TFIM}}$ governs the free (rotating-frame) evolution of the $N$ register spins in the top row of the array. The $\lceil N/2\rceil$ spins in the bottom row encode a repetition-code qubit. This qubit acts as the control qubit for gates performed on the register, which have the effect of toggling the analog-simulation parameters. By initializing the register in a mixed state, the $2^N$ eigenenergies of $H_{\mathrm{TFIM}}$ can be extracted from measurements of the control qubit, requiring measurements of just one physical spin. Classical postprocessing of the qubit measurement outcomes can then be used to isolate the $N$ single-particle energies of the Kitaev chain obtained from $H_{\mathrm{TFIM}}$ via the Jordan--Wigner transformation.  }
    \label{fig:setup}
\end{figure}

In this work, we show how the full spectrum of the transverse-field Ising model (TFIM) can be extracted from measurements of a single spin by stroboscopically matching the output of a  model of mixed-state quantum computation known as deterministic quantum computing with one clean qubit (DQC1)~\cite{knill1998power}. In a DQC1 circuit, only the ``clean'' qubit starts in a state having nonzero polarization, while all others are in a maximally mixed state. A unitary gate is applied to the spins in the mixed state, conditioned on the state of the clean qubit, and only the clean qubit is measured. It is known that the output of the DQC1 model cannot be classically efficiently sampled (to within multiplicative error) unless the polynomial-time hierarchy collapses to
the second level~\cite{fujii2018impossibility}, and that DQC1 can be used to perform tasks for which there are no known polynomial-time classical algorithms, such as estimating Jones polynomials~\cite{shor2008estimating}. It can also be used to estimate the normalized trace of any $n$-qubit unitary having an efficient implementation in terms of quantum gates, with a number of measurements that scales polynomially in the inverse target accuracy, independent of the size of the unitary~\cite{knill1998power,datta2005entanglement}.

In terms of physical device layout, we consider a $2\times N$ array of quantum dots---a common geometry for state-of-the-art devices~\cite{jirovec2026many,dijkema2026simultaneous}. In contrast to other qubit modalities, analog simulation of the TFIM can be realized by taking advantage of interactions native to spin qubits. We show that full spectral information can be mapped onto a single spin playing the role of the ``clean'' qubit by periodically toggling the analog-simulation parameters using two-qubit controlled-Y (CY) gates conditioned on the state of this qubit. These CY gates, in combination with free analog simulation, can then be used to engineer the Floquet Hamiltonian governing the stroboscopic evolution of the spins in the array. At stroboscopic times, the coherence of the clean qubit is given by the normalized trace of an (effective) multiqubit controlled unitary implementing time evolution under the TFIM Hamiltonian, as in a DQC1 circuit. The measured time-domain qubit coherence can then be classically postprocessed to recover the $N$ single-particle energies of the Kitaev chain obtained from the original TFIM via a Jordan--Wigner transformation. The protocol presented here could therefore be used to map, using measurements of just one spin, the spectral signature of the crossover from the trivial to the topological regime of the Kitaev chain.

The remainder of this article is organized as follows: We begin by describing the physical setup associated with the quantum simulation, before turning our attention to the gate sequence needed to stroboscopically match the output of the DQC1 model.  We then show how the single-particle spectrum of the Kitaev chain can be extracted from measurements of the clean-qubit coherence. Finally, we discuss practical constraints on the total simulation time and size $N$ of the array.

\section{Setup}

During the simulation, the $N$ ``register'' spins in the top row of the $2\times N$ array (Fig.~\ref{fig:setup}) are allowed to evolve freely under the action of some Hamiltonian, taken to be the Hamiltonian $H_{\mathrm{TFIM}}$ of the transverse-field Ising model ($\hbar=1$),
\begin{equation}\label{H0}
    H_{\mathrm{TFIM}}=h\sum_{j=1}^N \tau_j^x+J\sum_{j=1}^{N-1} \tau_j^z\tau_{j+1}^z.
\end{equation}
Here, $\tau_j^{x,z}$ are Pauli operators acting on the $j^{\mathrm{th}}$ spin, $h$ is the strength of the transverse field, and $J$ is the strength of the exchange interaction between neighboring spins. Notably, $H_{\mathrm{TFIM}}$ has a $\mathbbm{Z}_2$ spin-flip symmetry described by 
\begin{equation}\label{Z2}
    \hat{P}^\dagger H_{\mathrm{TFIM}}\hat{P}=H_{\mathrm{TFIM}},
\end{equation}
where $\hat{P}=\prod_j(-i\tau_j^x)$ describes a rotation by angle $\pi$ of all spins about the $x$ axis.  This operation flips the sign of every $\tau_j^z$ while leaving $\tau_j^x$ invariant.

By applying continuous Rabi driving to the register spins, $H_{\mathrm{TFIM}}$ can be realized in a frame rotating at the drive frequency of each spin without the requirement of natively Ising exchange in the lab frame (Appendix~\ref{appendix:rot-frame}). More broadly, Rabi driving can be used to define dressed qubits having improved coherence times~\cite{laucht2017dressed,kuno2026robust} and for control of large spin arrays using global driving fields~\cite{seedhouse2021quantum,hansen2021pulse,vahapoglu2022coherent,hansen2024entangling}. In the present case, driving enables in-situ tunability of the (synthetic) transverse field $h$ and allows the spins' integrability-breaking Zeeman terms to be eliminated in the rotating frame. 

For $J>h$, the TFIM is in an ordered phase in which the ground state is given, for $h=0$, by a degenerate doublet of ferromagnetic states in which all spins are oriented parallel or antiparallel to the $z$ axis; for finite $h$, these states hybridize into a Greenberger-Horne-Zeilinger (GHZ) doublet separated in energy by a splitting that is exponentially suppressed in $N$. For $J<h$, by contrast, the TFIM is in a disordered paramagnetic state. The phase transition at $J=h$ is only well defined in the thermodynamic limit and is replaced by a smooth crossover for finite $N$.

The TFIM can be mapped onto the Kitaev chain~\cite{kitaev2001unpaired} via a Jordan--Wigner transformation in which $\tau_i^{x}=1-2c_i^\dagger c_i$ and $\tau_i^{z}=\prod_{j<i}\tau_j^{x}(c_i+c_i^\dagger)$, where $c_i, c_i^\dagger$ are fermionic operators satisfying the usual fermionic anticommutation relations $\{c_i,c_j\}=0$ and $\{c_i,c_j^\dagger\}=\delta_{ij}$. Under this mapping, the chemical potential $\mu$ in the Kitaev chain is given by $\mu=2h$, while the superconducting pairing $\Delta$ and hopping amplitude $t_\mathrm{h}$ are related to the exchange coupling, with the resulting Hamiltonian $H_{\mathrm{K}}$ given by
\begin{equation}\label{kitaev-hamiltonian} 
    H_{\mathrm{K}}=-2h\sum_j\left(c_j^\dagger c_j-\frac{1}{2}\right)+J\sum_{\langle j,k\rangle} (c_j^\dagger c_k+c_j^\dagger c_k^\dagger+\mathrm{h.c.}).
\end{equation}
Since the hopping and pairing terms have equal magnitude, $H_\mathrm{K}$ lies along the sweet line $\vert t_\mathrm{h}\vert=\vert\Delta\vert$ of the Kitaev phase diagram. The ordered phase of the TFIM ($J>h$) then corresponds to the topological phase of the Kitaev chain, while the disordered phase ($h>J$) corresponds to the trivial phase, in which all excitations are gapped. We must emphasize, however, that the Jordan--Wigner transformation is nonlocal: In the spin representation, the Majorana end modes correspond to string operators supported on the entire chain, and the near-degeneracy of the two lowest eigenstates in the topological phase of the Kitaev chain is that of a symmetry-broken GHZ-like doublet, protected by the $\mathbbm{Z}_2$ spin-flip symmetry rather than by fermionic parity and locality~\cite{greiter2014ising,tserkovnyak2011universal,backens2017emulating}. What is faithfully preserved under the mapping (and what the protocol presented here measures) is the spectrum, including the near-zero-energy signature of the topological regime of the Kitaev chain.

An analog simulation of dynamics under the Kitaev-chain Hamiltonian, realized via the mapping to the TFIM discussed above, can be combined with digital gates on the register spins designed to toggle between different free-evolution Hamiltonians. These gates are conditioned on the state of a ``clean'' control qubit that resides in the bottom row of the array (Fig.~\ref{fig:setup}). The control qubit is itself encoded in a repetition code of $\lceil N/2\rceil$ physical spins, where $\lceil \cdot \rceil$ is the ceiling function. The encoding procedure can be realized with the native connectivity of the dot array by initializing a single physical spin in state $\ket{\psi}$, and then encoding the state of the single spin into the corresponding logical state $\ket*{\bar{\psi}}$ of the repetition code using a series of CNOT gates between neighboring spins. The purpose of the repetition code is not to allow error correction of the control qubit, but rather to enable parallelized operations between the control qubit and $\lceil N/2\rceil$ register spins. This capability follows from the fact that any code-space-preserving operation conditioned on, e.g., the state $\ket{\downarrow}$ of a physical spin in the repetition code is logically equivalent to the same operation conditioned on the logical basis state $\ket*{\bar{\downarrow}}$ of the encoded qubit. Measurements of the control qubit can be realized by decoding the logical state $\ket*{\bar{\psi}}$, which has the effect of mapping the encoded information back onto the state $\ket{\psi}$ of a single physical spin. Hence, independent of $N$, only one physical spin needs to be measured.

\section{Stroboscopic DQC1}

The protocol, expressed as a quantum circuit, consists of four stages: preparation of the control qubit in state $\ket*{\bar{+}}\propto\ket*{\bar{\uparrow}}+\ket*{\bar{\downarrow}}$, followed by $n$ rounds of evolution, and finally, decoding and measurement of the control qubit.  We take the Hamiltonian evolution of all spins to be generated by 
\begin{align}\label{rot-frame-hamiltonian}
\begin{aligned}
    H(t)=\frac{\omega_\mathrm{q}}{2}\sigma_z+H_{\mathrm{TFIM}}+H_\mathrm{DD}(t)+\Pi(t)\otimes H_\mathrm{c}(t).
\end{aligned}
\end{align}
Here,  $\sigma_z=\ketbra*{\bar{\uparrow}}-\ketbra*{\bar{\downarrow}}$ is the Pauli-Z operator of the control qubit, with its energy splitting $\omega_\mathrm{q}$ given by the sum of the splittings of the $\lceil N/2\rceil$ spins involved in the repetition code. The first two terms in $H(t)$ therefore describe free evolution of the control qubit and register spins, respectively. The third term $H_{\mathrm{DD}}(t)$ is a Hamiltonian that generates X gates on all spins encoding the control qubit at times $\zeta_n=(n-1)T+T/2$, $n\geq 1$, and on all spins in the register at times $\zeta_n'=n T/2$, $n\geq 1$, where here, $T$ is a timescale that will eventually correspond to a Floquet period (discussed below). Since flipping the state of all physical spins encoding the control qubit simply flips $\ket*{\bar{\uparrow}}\leftrightarrow\ket*{\bar{\downarrow}}$, these X gates implement a Carr--Purcell--Meiboom--Gill (CPMG) dynamical decoupling sequence on the control qubit and a periodic dynamical decoupling (PDD) sequence on individual register spins. The use of PDD instead of CPMG for the register will be clarified subsequently. The symbol $\Pi(t)$ denotes a projector onto a basis state of the control qubit, given by $\Pi(t)=\ketbra*{\bar{\uparrow}}$ for times $t\in [0, \zeta_1)$ and $t\in I_{2n}=(\zeta_{2n},\zeta_{2n+1})$, and by $\Pi(t)=\ketbra*{\bar{\downarrow}}$ for $t\in I_{2n+1}$. This flip of $\Pi(t)$ at every X gate in the CPMG sequence ensures that the evolution of the control qubit due to its coupling to the register spins is \textit{not} refocused by the dynamical decoupling. Finally, $H_\mathrm{c}(t)$ generates pulses on the register spins conditioned on the state of the control qubit, with the fourth term in Eq.~\eqref{rot-frame-hamiltonian} therefore describing the effects of two-qubit gates. We show in Appendix~\ref{appendix:rot-frame} how these gates could be realized in the rotating frame.

We first transform $H(t)$ to a frame obtained through the unitary transformation 
\begin{align}\label{qubit-transformation}
    &U_{0}(t)=U_\mathrm{DD}(t)U_\mathrm{q}(t),\\
    &U_\mathrm{DD}(t)=\mathcal{T}e^{-i\int_0^tds \:H_\mathrm{DD}(s)},\\
    &U_\mathrm{q}(t)=\mathcal{T}e^{-i\frac{\omega_\mathrm{q}}{2}\int_0^tds\: \hat{\sigma}_z(s)},\label{u-delta}
\end{align}
where in Eq.~\eqref{u-delta}, $\hat{\sigma}_z(t)=U_\mathrm{DD}^\dagger(t)\sigma_z U_\mathrm{DD}(t)$ accounts for the effect of the CPMG sequence on the control qubit. Treating the dynamical-decoupling pulses as ideal and effectively instantaneous on the timescale of evolution due to $H_{\mathrm{TFIM}}$~\cite{viola1998dynamical,viola2003robust,facchi2004unification}, we can write $U_\mathrm{DD}(t)$ as
\begin{align}\label{pi-pulses}
    U_\mathrm{DD}(t)=(-i\sigma_x)^{m(t)}\hat{P}^{\ell (t)},
\end{align} 
where, as before, $\hat{P}=\prod_{j}(-i\tau_j^x)$ is a product of Pauli-X operators acting on all spins in the register, and where $m(t)=\sum_{n}\Theta(t-\zeta_n)\in\mathbbm{Z}$ [$\ell(t)=\sum_n\Theta(t-\zeta_n')\in\mathbbm{Z}$] is the number of X gates that have been applied to the control qubit (register) up to time $t$, with $\Theta(t)$ denoting the Heaviside function. Under this transformation, the original Hamiltonian $H(t)$ is given by
\begin{equation}\label{hamiltonian-pi-frame}
    \hat{H}(t)=\mathbbm{1}\otimes \hat{H}_{\mathrm{TFIM}}(t)+\ketbra*{\bar{\uparrow}}\otimes \hat{H}_\mathrm{c}(t),
\end{equation}
where $\hat{H}_{\mathrm{TFIM}}(t)=[\hat{P}^{\ell(t)}]^\dagger H_{\mathrm{TFIM}} \hat{P}^{\ell (t)}$, and where $\hat{H}_\mathrm{c}(t)$ is defined analogously.

We now go into a frame toggling with respect to the second term in Eq.~\eqref{hamiltonian-pi-frame}, obtained via the unitary transformation 
\begin{equation}\label{toggling-frame-unitary}
    U_{\mathrm{TF}}(t)=\ketbra*{\bar{\uparrow}}\otimes U_{\mathrm{c}}(t)+\ketbra*{\bar{\downarrow}}\otimes\mathbbm{1},
\end{equation}
where $U_\mathrm{c}(t)=\mathcal{T}e^{-i\int_0^tds\hat{H}_\mathrm{c}(s)}$. In this frame, the Hamiltonian reads
\begin{equation}\label{toggling-frame-hamiltonian}
    \tilde{H}(t)=\ketbra*{\bar{\downarrow}}\otimes \tilde{H}_\downarrow(t)+\ketbra*{\bar{\uparrow}}\otimes  \tilde{H}_\uparrow(t),
\end{equation}
where 
\begin{align}
    &\tilde{H}_\downarrow(t)=\hat{H}_{\mathrm{TFIM}}(t),\label{H_down}\\
    &\tilde{H}_\uparrow(t)=U_\mathrm{c}^\dagger(t)\hat{H}_{\mathrm{TFIM}}(t)U_\mathrm{c}(t).\label{H_up}
\end{align}
Since $H_{\mathrm{TFIM}}$ has a $\mathbbm{Z}_2$ symmetry generated by $\hat{P}$ [Eq.~\eqref{Z2}], we have $\hat{H}_{\mathrm{TFIM}}(t)=H_{\mathrm{TFIM}}$ in Eqs.~\eqref{H_down}-\eqref{H_up}. This symmetry is what ensures that our analog simulation of the TFIM remains compatible with dynamical decoupling of the individual register spins, which would otherwise average out any terms in the free-evolution Hamiltonian that do not commute with $\hat{P}$. 

Similar to what was done for $U_\mathrm{DD}(t)$, the unitary $U_\mathrm{c}(t)$ describing the pulses applied to the register spins can be decomposed into a product of gates $\hat{G}_j$ applied at times $t_j$ under the assumption that the pulses implementing these gates are fast on the timescale of the free evolution generated by $H_{\mathrm{TFIM}}$. If the Hamiltonian $H_{\mathrm{DD}}(t)$ implementing $\pi$ pulses on the register does not overlap in time with the control Hamiltonian $H_{\mathrm{c}}(t)$, then 
\begin{equation}\label{control-unitary}
    U_\mathrm{c}(t)=\hat{G}_j\dotsm \hat{G}_1,\quad t_{j}<t<t_{j+1},
\end{equation}
where $\hat{G}_j=[\hat{P}^{\ell(t_j)}]^\dagger G_j\hat{P}^{\ell(t_j)}$ with $G_j$ denoting a gate implemented by $H_{\mathrm{c}}(t)$. The explicit $G_j$ are given later in this section.

The quantity that will eventually encode information about $H_{\mathrm{TFIM}}$ is the average coherence of the control qubit, given by
\begin{equation}\label{coherence}
    \langle \sigma_+\rangle_{t}=\mathrm{Tr}\{\sigma_+\rho(t)\},
\end{equation}
where $\sigma_+=\ketbra*{\bar{\uparrow}}{\bar{\downarrow}}$ and  $\rho(t)=U(t)\rho(0)U^\dagger(t)$ with $U(t)=\mathcal{T}e^{-i\int_0^t ds\:H(s)}$.  The qubit coherence can be measured straightforwardly by first decoding the repetition code. This can be achieved by inverting the encoding circuit, which has the effect of mapping the state of the encoded control qubit onto the state of a single spin. All other spins involved in the repetition code end up in a product state decoupled from the rest of the system and may therefore be traced over with no impact on the expectation value given in Eq.~\eqref{coherence}.  Although $H_{\mathrm{TFIM}}$ is realized in a frame rotating at the Rabi-drive frequency, control-qubit observables are unaffected by this transformation since it acts trivially on the Hilbert space of the control qubit. The decoded control-qubit state, and hence the coherence in Eq.~\eqref{coherence}, can therefore be measured directly in the lab frame without any additional transformation of the measurement basis.

Since $U(t)=U_0(t)U_{\mathrm{TF}}(t)\tilde{U}(t)$, Eq.~\eqref{coherence} can equivalently be written as 
\begin{equation}
    \langle \sigma_+\rangle_{t}=\mathrm{Tr}\{\hat{\sigma}_+(t)U_{\mathrm{TF}}(t)\tilde{\rho}(t)U_{\mathrm{TF}}^\dagger(t)\},
\end{equation}
where $\hat{\sigma}_+(t)=U_0^\dagger(t)\sigma_+ U_0(t)$ and $\tilde{\rho}(t)=\tilde{U}(t)\rho(0)\tilde{U}^\dagger(t)$.  Using Eq.~\eqref{pi-pulses}, the time-dependent spin-raising operator $\hat{\sigma}_+(t)$ can be written for times $t\in I_n$ as
\begin{equation}\label{qubit-operator}
    \hat{\sigma}_+(t)=\begin{cases}
        e^{i\phi(t)}\sigma_+,\quad &n\text{ even}\\
        e^{-i\phi(t)}\sigma_-,\quad &n\text{ odd}
    \end{cases},
\end{equation}
where $\phi(t)=\omega_\mathrm{q}\int_0^tdt's(t')$ with $s(t)=(-1)^{m(t)}$.  As a result of the CPMG sequence applied to the control qubit, $\phi(nT)=0$.

\begin{figure}
    \centering
    \includegraphics[width=\linewidth]{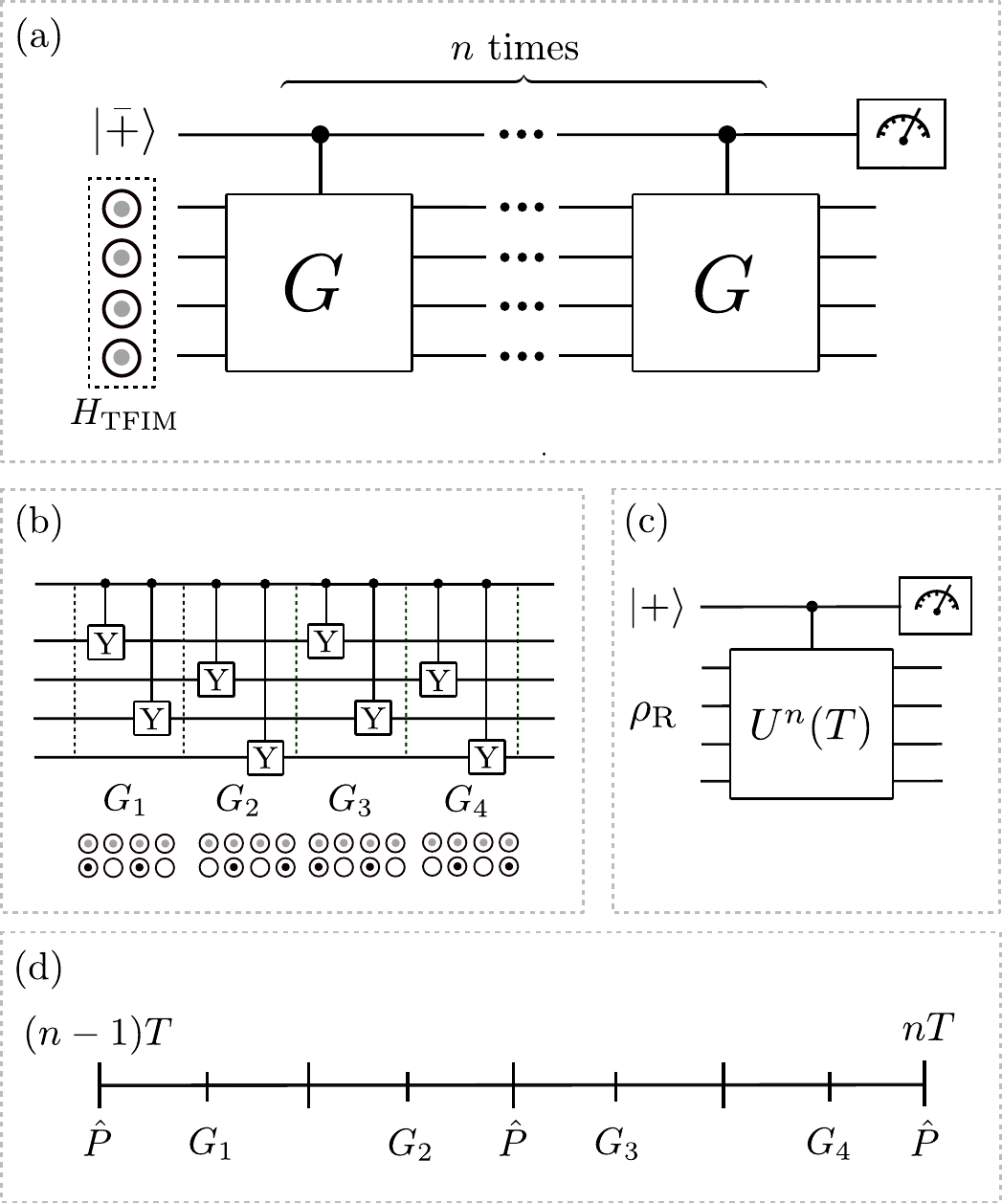}
    \caption{(a) Circuit representation of the quantum simulation. The register spins evolve freely under $H_{\mathrm{TFIM}}$, with information mapped onto the control qubit through a periodic sequence of two-qubit gates acting on the control qubit and  register. (b) Gate sequence comprising the controlled-$G$ gate for $N=4$. At each time step $t_j$ ($j=1,\dots,4$), CY gates are applied to the $N/2$ register spins indexed by even or odd integers, with the corresponding charge configurations in the quantum-dot array shown below as filled or empty circles. The gates in each $G_j$ can be applied in parallel due to the repetition encoding used for the control qubit. Gates implementing dynamical decoupling of the control qubit and register are not illustrated. (c) For register spins initially prepared in state $\rho_\mathrm{R}$ and evolving freely under $H_{\mathrm{TFIM}}$, the circuit in (a) is equivalent to the circuit shown here for $U(T)=e^{-i H_{\mathrm{TFIM}}T}$. When $\rho_{\mathrm{R}}$ is taken to be maximally mixed, this circuit represents an instance of the one-clean-qubit (DQC1) model. (d) Timing of the gate sequence $G_1,\dots,G_4$ within the Floquet period $T$, including the gates $\hat{P}=\prod_j(-i\tau_j^x)$ implementing the PDD sequence on the register spins.}
    \label{fig:circuits}
\end{figure}

We take the initial state $\rho(0)$ of the control qubit and $N$-spin register to be a product state of the form 
\begin{equation}
    \rho(0)=\ketbra*{\bar{+}}\otimes \rho_{\mathrm{R}}(0),
\end{equation}
where $\rho_{\mathrm{R}}(0)=(\mathbbm{1}/2)^{\otimes N}$ is a maximally mixed state~\cite{knill1998power}.  A maximally mixed state is straightforward to prepare when the relaxation time $T_1$ of the register spins far exceeds their dephasing time $T_2^*$, as is typical for spin qubits. Starting from the all-ground-state configuration, a uniform superposition of computational basis states is prepared by applying a Hadamard  to each spin. The system is then allowed to decohere for a time $t_{\mathrm{prep}}$ satisfying $T_2^*\ll t_{\mathrm{prep}}\ll T_1$, during which coherences between basis states decay while populations remain unchanged (to a good approximation). This yields the required maximally mixed state. For spin qubits in gate-defined quantum dots, relaxation times typically reach the order of seconds, while dephasing times typically fall in the range of tens of microseconds in the absence of dynamical decoupling~\cite{stano2022review}.

The spectrum of $H_{\mathrm{TFIM}}$ can then be extracted from measurements of the control qubit by choosing an appropriate control sequence $U_\mathrm{c}(t)$ [Eq.~\eqref{control-unitary}], as we now show. First, we impose the requirement that $U_{\mathrm{c}}(nT)=\mathbbm{1}$, which can be satisfied by ensuring that every gate $\hat{G}_j$ appearing in Eq.~\eqref{control-unitary} over the subinterval $[nT,(n+1)T]$ is inverted by the end of that subinterval. With this condition satisfied, the unitary $U_{\mathrm{TF}}(t)$ generating the toggling-frame transformation is equal to the identity at stroboscopic times $t=nT$, $U_{\mathrm{TF}}(nT)=\mathbbm{1}$ [cf.~Eq.~\eqref{toggling-frame-unitary}]. For alternating even and odd $n$, the average qubit coherence $\langle\sigma_+\rangle_t$ at $t=nT$ is then given by 
\begin{align}\label{sigma-plus-even-odd}
    \langle\sigma_+\rangle_{nT}=\frac{1}{2}\mathcal{K}^n
        \mathrm{Tr}_{\mathrm{R}}\{\tilde{U}_\uparrow^\dagger(nT)\tilde{U}_\downarrow(nT)\rho_\mathrm{R}(0)\},
\end{align}
where $\mathcal{K}$ satisfies $\mathcal{K}z=z^*$ for $z\in\mathbbm{C}$, $\tilde{U}_{\alpha}(t)=\langle\bar{\alpha}\vert \tilde{U}(t)\vert\bar{\alpha}\rangle$ with $\tilde{U}(t)=\mathcal{T}e^{-i\int_0^tds\tilde{H}(s)}$, and where $\mathrm{Tr}_{\mathrm{R}}\{\cdot\}$ denotes a partial trace over the state of the register. Since the cases of even versus odd $n$ in Eq.~\eqref{sigma-plus-even-odd} are related through complex conjugation, the coherence $\langle\sigma_+\rangle_{nT}$ could be reconstructed through classical postprocessing of the qubit measurement outcomes (since $\sigma_\pm\propto\sigma_x\pm i\sigma_y$).

In addition to the requirement that $U_\mathrm{c}(nT)=\mathbbm{1}$, we also assume that $U_\mathrm{c}(t+T)=U_\mathrm{c}(t)$. The periodic sequence of pulses comprising $U_\mathrm{c}(t)$ [cf.~Eq.~\eqref{control-unitary}] can then be used to engineer the Floquet Hamiltonian $H_{\mathrm{F},\uparrow}$ describing the stroboscopic evolution of the register spins conditioned on the state of the control qubit being $\ket*{\bar{\uparrow}}$:
\begin{equation}
    \tilde{U}_\uparrow(nT)=e^{-i H_{\mathrm{F},\uparrow}nT}.
\end{equation}
Formally, the Floquet Hamiltonian $H_{\mathrm{F},\uparrow}$ can be evaluated through a Floquet-Magnus expansion~\cite{blanes2009magnus} as
\begin{equation}\label{floquet-magnus}
    H_{\mathrm{F},\uparrow}=\sum_{k=0}^\infty T^k\Lambda_k,
\end{equation}
where $\Lambda_k$ is an operator involving $k$ nested commutators.
However, to leading order in this expansion [Eq.~\eqref{floquet-magnus}], $H_{\mathrm{F},\uparrow}$ is given simply by the period-averaged Hamiltonian $\bar{H}_\uparrow\equiv \Lambda_0$:
\begin{align}
    &H_{\mathrm{F},\uparrow}=\bar{H}_\uparrow +O(\Vert H_\mathrm{TFIM}\Vert^2 T),\label{magnus-expansion}\\
    &\bar{H}_\uparrow=\frac{1}{T}\int_0^T dt\:\tilde{H}_\uparrow(t).
\end{align}
For bounded, finite-dimensional systems, a sufficient condition for convergence of the Floquet-Magnus expansion is $\int_0^T dt\,\Vert\tilde{H}_\uparrow(t)\Vert<\pi$~\cite{blanes2009magnus}. With $\Vert H_{\mathrm{TFIM}}\Vert=E_{\mathrm{max}}$ given by the eigenvalue of $H_{\mathrm{TFIM}}$ having the largest magnitude, the expansion then converges when $E_{\mathrm{max}}T/\pi<1$.

Two-qubit gates between the control qubit and register spins can be realized by turning on the exchange interaction between spins in the top and bottom rows of the array. Since the control qubit is encoded in a repetition code, a two-qubit gate conditioned on the state $\ket{\uparrow}$ of a spin involved in the encoding is logically equivalent to a two-qubit gate conditioned on $\ket*{\bar{\uparrow}}$. The choice of encoding in combination with the spatial layout of the array therefore allows two-qubit gates to be executed in parallel with a one-to-many connectivity [Fig.~\ref{fig:circuits}(a)]. This, in turn, allows $\bar{H}_\uparrow$ to be engineered and ultimately canceled with a gate sequence having a duration independent of $N$. With $\bar{H}_{\mathrm{F},\uparrow}\simeq \bar{H}_\uparrow=0$, the qubit coherence is then directly controlled by the eigenenergies of the TFIM, as will be clarified later in this section.

The cancellation of $\bar{H}_\uparrow$ can be realized by flipping the spins in the register, conditioned on the state of the control qubit, so that $h$ and $J$ both change sign for half the Floquet period $T$. Due to the two-spin interactions in $H_{\mathrm{TFIM}}$, changing the sign of $J$ requires that neighboring spins be addressed at different times. At times $t_j$, where $j=1,\dots,4$, the $\lceil N/2\rceil$ spins encoding the control qubit each act as the physical control qubit for at most one two-qubit gate acting on a register spin (exactly one in the case where $N$ is even). The specific values of $t_j$ will be specified later in this section. As discussed previously, these operations are logically equivalent to a set of simultaneous two-qubit gates between the control qubit and register spins. We denote by $G_j$ the operations applied to the register conditioned on the state $\ket*{\bar{\uparrow}}$ of the control qubit in the rotating frame [Eq.~\eqref{hamiltonian-pi-frame}]. These gates $G_j$ appeared previously in the toggling-frame description of the Hamiltonian  [Eqs.~\eqref{toggling-frame-hamiltonian}--\eqref{control-unitary}], and concretely, for $j$ even (odd), all gates in $G_j$ act on spins indexed by even (odd) integers only. The gates themselves are given by
\begin{align}
\begin{aligned}\label{control-pulses}
    &G_1, G_3=\prod_{i\:\mathrm{odd}}Y_{i},\\ 
    &G_2,G_4=\prod_{i\:\mathrm{even}}Y_{i},
\end{aligned}
\end{align}
where $Y_i=-i\tau_i^y$ denotes a Y gate applied to the $i^{\mathrm{th}}$ register spin [Fig.~\ref{fig:circuits}(b)]. Due to the X gates on the register at the midpoint $\zeta_{2n+1}'=(n+1/2)T$ of the Floquet interval, we then have
\begin{align}\label{uc(T)}
    U_\mathrm{c}(T)&=(\hat{P}^\dagger G_4\hat{P})(\hat{P}^\dagger G_3 \hat{P})G_2G_1=\mathbbm{1}. 
\end{align}
This ensures that $U_\mathrm{c}(nT)=\mathbbm{1}$ as required. Note that the  X gates at times $\zeta_{2n}'=nT$, which differentiate the PDD sequence applied to the register spins from the CPMG sequence applied to the control qubit, are needed to ensure that the next gate (belonging to the next Floquet cycle) is $G_1$ rather than $\hat{P}^\dagger G_1\hat{P}$ since the latter would invalidate the assumption made above that $U_{\mathrm{c}}(t)$ is periodic with period $T$. We emphasize once more that the ability to write $U_{\mathrm{DD}}(t)$ and $U_{\mathrm{c}}(t)$ in terms of individual gates, as done in Eqs.~\eqref{pi-pulses} and \eqref{uc(T)}, proceeds from the assumption made above that the dynamics due to $H_{\mathrm{TFIM}}$ are negligible on the timescale over which gates are performed. This amounts to setting $H_{\mathrm{TFIM}}= 0$ while $H_{\mathrm{c}}(t)\neq 0$ or $H_{\mathrm{DD}}(t)\neq 0$, consistent with setting $\tilde{H}_{\uparrow,\downarrow}(t)= 0$ over the duration of a gate.

During the free-evolution portion of the Floquet cycle (i.e.,~when no gate is being performed), the Hamiltonian $\tilde{H}_\uparrow(t)$ will consequently toggle between four different Hamiltonians $H_j=G_{\mathrm{tot},j}^\dagger H_{\mathrm{TFIM}}G_{\mathrm{tot},j}$, $j=1,\dots, 4$, where $G_{\mathrm{tot},j}=\prod_{j'\leq j}G_{j'}$. Explicitly, these Hamiltonians are given by
\begin{align}
    &H_1=+h\sum_{\ell=1}^N(-1)^\ell\tau_\ell^x-J\sum_{\ell=1}^{N-1}\tau_\ell^z\tau_{\ell+1}^z,\\
    &H_2=-h\sum_{\ell=1}^N\tau_\ell^x+J\sum_{\ell=1}^{N-1}\tau_\ell^z\tau_{\ell+1}^z,\\
    &H_3=-h\sum_{\ell=1}^N(-1)^\ell\tau_\ell^x-J\sum_{\ell=1}^{N-1}\tau_\ell^z\tau_{\ell+1}^z,\\
    &H_4=+h\sum_{\ell=1}^N\tau_\ell^x+J\sum_{\ell=1}^{N-1}\tau_\ell^z\tau_{\ell+1}^z.
\end{align}
Due to the $\mathbb{Z}_2$ symmetry of the TFIM, the Hamiltonian $\tilde{H}_\downarrow(t)$ is simply given by $H_{\mathrm{TFIM}}$ as remarked previously.

With this gate sequence, $\tilde{H}_\downarrow(t)$ and $\tilde{H}_\uparrow(t)$ both retain the structure of a TFIM at all times $t$, up to a site-dependent staggering of the transverse field in the case of $\tilde{H}_\uparrow(t)$. The times $t_j$ of the CY gates should be chosen so that each Hamiltonian above drives evolution of the register, conditioned on the control qubit being in state $\ket*{\bar{\uparrow}}$, for the same amount of time $\Delta T$. For a control qubit in state $\ket*{\bar{\uparrow}}$, the stroboscopic evolution of the register is then effectively frozen:
\begin{equation}
    \bar{H}_\uparrow=\frac{1}{T}\sum_{j=1}^{4}H_j\Delta T =0.\label{frozen}
\end{equation}
This may be understood as a form of Hamiltonian engineering via dynamical decoupling~\cite{rotteler2006equivalence,evert2025syncopated,brown2025efficient,nguyen2026color}. The cancellation in Eq.~\eqref{frozen} also holds in the presence of site-dependence in the strength of the transverse field and exchange ($h,J\mapsto h_\ell,J_{\ell,\ell+1}$), making it robust to control imperfections in the engineering of the TFIM.

In the limit of infinitesimally fast gates, Eq.~\eqref{frozen} can be realized by choosing $t_j/T=(2j-1)/8$ (mod 1), in which case $\Delta T=T/4$ [Fig.~\ref{fig:circuits}(d)]. For finite gate times, however, the $t_j$ should be spaced unequally to compensate for the dynamical-decoupling gates applied during certain free-evolution intervals. Under the assumption that the shuttling time is negligible compared to the time needed for single- and two-qubit gates, the duration $T$ of the Floquet cycle is then given by $T= 4\Delta T +T_{\mathrm{g}}$, independent of $N$, where $T_{\mathrm{g}}$ is the time needed to implement the gates. The timescale associated with repositioning the control qubit in the quantum-dot array is controlled by the need for adiabatic charge transfer between dots. For bucket-brigade shuttling, orbital splittings typically translate to a timescale of a few nanoseconds for transporting spins between sites~\cite{wang2024operating,unseld2025baseband}.

Along similar lines, it may be determined that the period-averaged Hamiltonian for a control qubit in state $\ket*{\bar{\downarrow}}$ is given by
\begin{equation}
    \bar{H}_\downarrow= \lambda H_{\mathrm{TFIM}},\quad\lambda=\left(1-\frac{T_{\mathrm{g}}}{T}\right),
\end{equation}
where the renormalization factor $\lambda$ accounts for the finite duration of the gates applied to the register. Neglecting higher-order terms in the Floquet-Magnus expansion, the average coherence of the control qubit at stroboscopic times $t=nT$ [cf.~Eq.~\eqref{sigma-plus-even-odd}] is then given by
\begin{equation}\label{general-sigma-plus}
    \langle\sigma_+\rangle_{nT}\simeq\frac{1}{2}\mathcal{K}^n\mathrm{Tr}_{\mathrm{R}}\{\tilde{U}_\downarrow(nT)\rho_{\mathrm{R}}(0)\}.
\end{equation}
Under the periodic sequence of CY gates specified above, the expectation values of X- and Y-basis measurements of the control qubit are therefore equivalent to those obtained for a circuit in which a controlled multiqubit gate $U(T)=e^{-i\lambda H_{\mathrm{TFIM}}T}$ is applied to all spins in the register [Fig.~\ref{fig:circuits}(c)]. When $\rho_{\mathrm{R}}$ is taken to be a maximally mixed state, Fig.~\ref{fig:circuits}(c) illustrates a DQC1 circuit~\cite{knill1998power} in which the controlled unitary corresponds to free evolution under the TFIM Hamiltonian.  With such a maximally mixed initial condition, Eq.~\eqref{general-sigma-plus} gives
\begin{align}
    \mathcal{K}^n\langle\sigma_+\rangle_{nT}\simeq\frac{1}{2^{N+1}}\sum_{j=0}^{2^N-1}e^{-iE_jnT},\label{DQC1}
\end{align}
where $E_j$ is an eigenvalue of $\lambda H_{\mathrm{TFIM}}$. Estimating the normalized trace of a unitary using a system with limited coherence (as realized in high-temperature NMR) is the task for which DQC1 was originally introduced~\cite{knill1998power}. With the protocol considered here, however, the control qubit encodes the normalized trace of $\tilde{U}_\downarrow$ even though $\tilde{U}_\downarrow$ is not directly being implemented as a multiqubit controlled operation, which would be challenging to realize without the use of Trotterization and three-qubit gates. 

Following from Eq.~\eqref{DQC1}, the main error due to finite-duration gates is a systematic reduction of the measured eigenenergies by a factor $\lambda$ (relative to the eigenenergies of $H_{\mathrm{TFIM}}$ itself). However, since this shift is known, it could easily be compensated if desired. An additional, arguably more important implication of this renormalization is that longer measurement times may be required to resolve individual energies, since energy differences are similarly reduced by a factor of $\lambda$. Requirements and constraints on the total measurement time are discussed in Sec.~\ref{sec:parameters}. 

Obtaining an estimate of $\langle\sigma_+\rangle_{nT}$ at a fixed number $n$ of Floquet cycles requires many measurements: The number $N_m$ of measurements needed to estimate the expectation value with accuracy $\delta$ scales like $N_m\sim \ln(1/P_e)/\delta^2$~\cite{datta2005entanglement}, where $P_e$ is the probability that the estimate is further from the true value by an amount exceeding $\delta$. Notably, $N_m$ is independent of the size $2^N$ of the Hilbert space. The polynomial scaling of $N_m\sim \delta^{-2}$, independent of $N$, underpins the assertion that DQC1  can be used to perform certain tasks efficiently (including trace estimation~\cite{knill1998power}) for which no polynomial-time classical algorithms are known. For the application considered in this work, the trace can be computed classically efficiently, so the strength of the DQC1 setup is its efficiency in terms of the number of spins that must be measured to extract the spectrum. This is especially relevant for near-term spin-qubit architectures, in which the number of charge sensors (used for qubit readout) is typically limited.

\section{Extracting the single-particle spectrum of the Kitaev chain}

Although $H_{\mathrm{TFIM}}$ acts on a Hilbert space of dimension $2^N$ and therefore has  $2^N$ eigenvalues,  the Kitaev chain admits a single-particle description in terms of Bogoliubov quasiparticle modes. This single-particle spectrum can be recovered through further classical postprocessing of the qubit measurement outcomes: Under the Jordan--Wigner transformation, $H_{\mathrm{TFIM}}$ maps onto the Hamiltonian $H_{\mathrm{K}}$ of the Kitaev chain given in Eq.~\eqref{kitaev-hamiltonian}, which can be written in the Nambu basis as
\begin{equation}
    H_{\mathrm{K}}=\frac{1}{2}\Psi^\dagger H_{\mathrm{BdG}}\Psi+\mathrm{const}.,
\end{equation}
where $\Psi=(\bm{c},\bm{c}^\dagger)^\top$ is the Nambu spinor and $H_{\mathrm{BdG}}$ is the Bogoliubov--de~Gennes (BdG) Hamiltonian whose $2N$ eigenvalues $\pm \epsilon_\alpha$ define the quasiparticle (single-particle) spectrum of the Kitaev chain. The eigenenergies $E_j$ of the TFIM Hamiltonian can therefore be written as sums of single-particle energies $\epsilon_\alpha$ of the form 
\begin{equation}\label{e-j}
    E_j=\sum_{\alpha=0}^{N-1}\epsilon_\alpha\left(n_\alpha-\frac{1}{2}\right),
\end{equation}
where $n_\alpha\in\{0,1\}$ is the occupation number of a quasimode, and where it then becomes natural to associate the index $j=0,\dots,2^N-1$ with the integer whose binary digits are $n_{N-1}\dots n_1n_{0}$. Using Eq.~\eqref{e-j}, the sum over $j$ in Eq.~\eqref{DQC1} can then be evaluated in terms of the single-particle spectrum only, giving
\begin{equation}\label{sum-to-product}
    \sum_j e^{-i E_jnT}=\prod_\alpha 2\cos{\left(\frac{\epsilon_\alpha nT}{2}\right)}.
\end{equation}

The expectation value of $\sigma_y$ is ideally $\langle\sigma_y\rangle_{nT}=0$ for all $n$, following from Eqs.~\eqref{DQC1} and \eqref{sum-to-product}. As a result, only $X$-basis measurements are needed to estimate $\langle \sigma_+\rangle_{nT}=(1/2)\langle\sigma_x\rangle_{nT}$, and the dependence on $\mathcal{K}^n$ becomes trivial. The product of cosines in Eq.~\eqref{sum-to-product} can be turned back into a sum by taking the logarithm of the control-qubit expectation value. We therefore define
\begin{equation}\label{S(nT)-def}
    S(nT)=\ln{\vert \langle\sigma_x\rangle_{nT}\vert}+N\ln{2},
\end{equation}
where the second term compensates for what would otherwise be a large DC offset due to the normalization of the trace in Eq.~\eqref{DQC1}. Using Eq.~\eqref{sum-to-product}, $S(nT)$ can be evaluated as
\begin{align}
    S(nT)&\simeq\sum_\alpha\ln{\bigg\vert 2\cos{\left(\frac{\epsilon_\alpha nT}{2}\right)}\bigg\vert}\label{s(nt)-1}\\
    &=\sum_\alpha\sum_{\ell=1}^\infty\frac{(-1)^{\ell-1}}{\ell}\cos{(\ell n\epsilon_\alpha T)},
\end{align}
where the approximation in Eq.~\eqref{s(nt)-1} was inherited from Eq.~\eqref{DQC1}, and where the second equality is valid provided $\epsilon_\alpha nT\neq (2k+1)\pi$ for $k\in\mathbbm{Z}$. Since this condition defines a measure-zero set of values of $\epsilon_\alpha T$, it is not expected to be satisfied for generic parameters and can likely be neglected in the context of a realistic experiment.

\begin{figure*}
    \centering
    \includegraphics[width=\linewidth]{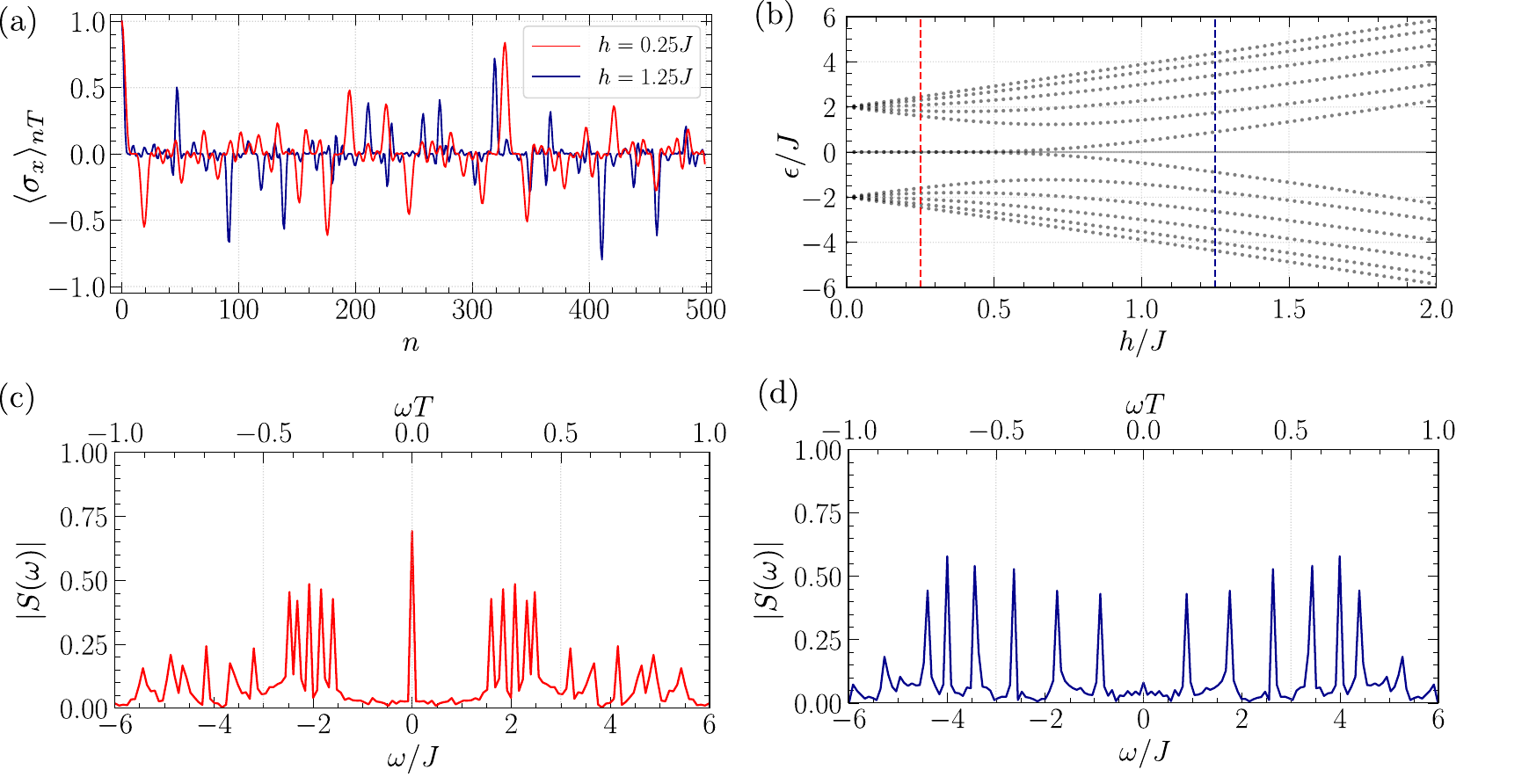}
    \caption{Simulations for $N=6$: (a) Expectation value of $\sigma_x$ at stroboscopic times $t=nT$ for $h=0.25J$ and $h=1.25J$, averaged over $10^3$ noise realizations. The simulations were performed with $J=2\pi\times 50$ kHz, $T_2^*=30$ $\mu$s, and $T=500$ ns, corresponding to a total simulation time of $T_{\mathrm{tot}}=250$ $\mu$s. (b) Single-particle spectrum of the Kitaev chain as a function of $h/J$, illustrating the splitting of the nearly degenerate ground state as the strength $h$ of the transverse field (equivalently, chemical potential) is increased.  The values of $h/J$ used in (c) and (d) are indicated by vertical dashed red and blue lines, respectively. (c) Spectrum $\vert S(\omega)\vert$ of the postprocessed qubit measurement outcomes for $h/J=0.25$, corresponding to the topological regime of the Kitaev chain, as a function of both $\omega/J$ (bottom axis) and $\omega T$ (top axis). The spectrum was calculated using a fast Fourier transform of the simulated coherence data and has been normalized so that $\vert S(\omega=0)\vert=1$ for $S(t)=1$. (d) Spectrum $\vert S(\omega)\vert$ for $h/J=1.25$, corresponding to the trivial regime. }
    \label{fig:spectrum}
\end{figure*}

The energies $\epsilon_\alpha$ can then be extracted by taking the discrete Fourier transform of $S(nT)$. The resulting spectrum $S(\omega)$ will feature dominant peaks at $\pm \epsilon_\alpha$, with higher harmonics suppressed in comparison. This postprocessing of the control-qubit measurement outcomes allows the single-particle spectrum of the Kitaev chain to be extracted, and by varying $h/J$, the crossover from the topological to the trivial regime can be mapped out. 

\subsection{Simulation}

We perform a numerical simulation of this experiment by integrating the time-dependent Schr\"odinger equation for evolution under $\tilde{H}_{\uparrow,\downarrow}(t)$. To model dephasing of the register spins, we include additional terms $\eta_j(t)\tau_j^z/2$ in $H(t)$ modeling classical, stochastic variation of the spin splittings. The time-domain trajectories $\eta_j(t)$ are related to the spectral densities $\Phi_j(\omega)=\int dte^{i\omega t}\langle\eta_j(t)\eta_j(0)\rangle$ through the relation $\langle \vert \eta_j(\omega)\vert^2\rangle=2\pi \Phi_j(\omega)$, where $\langle \cdot \rangle$ denotes an average over noise realizations. For a given realization, we therefore take $\eta_j(t)$ to be given by the inverse Fourier transform of $\eta_j(\omega)=[\pi \Phi_j(\omega)]^{1/2}[x_j(\omega)+i y_j(\omega)]$, where $x_j$ and $y_j$ are independent, zero-mean Gaussian random variables with unit variance. The spectral density was taken to be $1/f$, $\Phi_j(\omega)\propto \omega^{-1}$, with its overall scale set by the spin's dephasing time $T_2^*$  (taken to be the same for all spins) according to $(T_2^*)^{-1}=\pi^{-1}\int d\omega \Phi_j(\omega)$. The lowest frequency relevant to a single experimental shot is set by the total duration $N_{\mathrm{max}}T$ of the experiment (with $N_{\mathrm{max}}$ corresponding to the total number of Floquet cycles), thereby limiting the total noise power via an effective infrared cutoff $\propto (N_{\mathrm{max}}T)^{-1}$. Gates on the register spins were taken to be ideal; we estimate the effects of gate errors in Sec.~\ref{sec:parameters}.

Numerical integration of the Schr\"odinger equation allows for evaluation of $\mathrm{Tr}_{\mathrm{R}}\{\tilde{U}_\uparrow^\dagger(nT)\tilde{U}_\downarrow(nT)\}$, accounting for both dephasing of the register and higher-order terms in the Floquet-Magnus expansion. The real part of this quantity sets the expectation value associated with an $X$-basis measurement of the control qubit [Fig.~\ref{fig:spectrum}(a)].  For different values of $h/J$, the single-particle spectrum of the Kitaev chain either features a pair of states close to zero energy or a finite bulk gap [Fig.~\ref{fig:spectrum}(b)]. From the simulated coherence data, these single-particle energies can be isolated by classically postprocessing the measured qubit coherence to obtain $S(nT)$ [Eq.~\eqref{S(nT)-def}], and then taking the discrete Fourier transform of $S(nT)$ [Figs.~\ref{fig:spectrum}(c,d)]. 

Since the number of register spins is finite, there is no sharp phase transition at $h=J$ as would occur in the thermodynamic limit. Furthermore, for nonzero $h\ll J$, the two states near zero energy are not perfectly degenerate. Instead, they are separated by a finite splitting $\Delta E/J\sim e^{-N/\xi}$, where $\xi^{-1}=\ln(J/h)$~\cite{kitaev2001unpaired}. For the choice $h=0.25J$ used in Fig.~\ref{fig:spectrum}, the splitting found from exact diagonalization is given by $\Delta E/J\approx 4.6\times 10^{-4}$. Spectrally resolving the oscillations due to this finite splitting would therefore require measurement times in excess of $\approx 40$ ms.

Dephasing of the control qubit was not simulated explicitly in Fig.~\ref{fig:spectrum}, but its effects can be calculated straightforwardly for the case of Gaussian $1/f$ noise. Since the state of the control qubit is preserved by its interaction with the register spins [cf.~Eq.~\eqref{toggling-frame-hamiltonian}], random variations $\delta\omega_\mathrm{q}(t)$ in the qubit splitting will lead to an extra phase in Eq.~\eqref{qubit-operator} given by $\delta\phi(t)=\int_0^t dt'\delta\omega_\mathrm{q}(t')s(t')$. Under the assumption that $\delta\omega_\mathrm{q}$ is stationary and Gaussian, averaging the qubit coherence over noise realizations recovers Eq.~\eqref{sigma-plus-even-odd}, up to a reduction in visibility given by $e^{-\chi(nT)}$, where $\chi(t)=(2\pi)^{-1}\int d\omega \Phi(\omega)F(\omega,t)$. Here, $\Phi(\omega)$ is the spectral density of $\delta\omega_\mathrm{q}(t)$, and $F(\omega,t)=\vert \int_0^t dt' e^{i\omega t'} s(t')\vert^2$ is a filter function  set by the pulse sequence $s(t)$. We assume $1/f$ noise and consequently take $\Phi(\omega)=\nu/\vert \omega\vert$. For $s(t)$ corresponding to a CPMG sequence with pulse spacing $T$, we then have~\cite{cywinski2008how}
\begin{equation}\label{chi(nT)}
    \chi(t=nT)\simeq \frac{C}{2\pi}\frac{\nu(nT)^2}{n},
\end{equation}
where $C=7\zeta_{\mathrm{R}}(3)/\pi^2\simeq 0.85$, with $\zeta_{\mathrm{R}}(z)$ denoting the Riemann zeta function.  Under the assumption that all $\lceil N/2\rceil$ spins encoding the control qubit are subject to independent Gaussian noise, the effective dephasing time of the control qubit is given by $T_2^*/\sqrt{\lceil N/2\rceil}$, where $T_2^*$ is the dephasing time of an individual spin. The noise amplitude $\nu$ is therefore related to the single-spin $T_2^*$ through the relation~\cite{makhlin2004dissipative, fehse2023generalized}
\begin{equation}
    \nu\simeq \frac{2\pi \lceil N/2\rceil}{(T_2^*)^2\ln{\left(\frac{\sqrt{\lceil N/2\rceil}}{\omega_{\mathrm{IR}}T_2^*}\right)}},
\end{equation}
where the infrared cutoff $\omega_{\mathrm{IR}}=2\pi/T_m$ is controlled by the total time $T_m$ over which the noise is probed. Taking $T=500$ ns, $T_2^*=30$ $\mu$s, and $T_m=1$ hour, the expected reduction in $\langle\sigma_x\rangle_{nT}$ due to dephasing after $n=500$ Floquet cycles is $e^{-\chi}\approx 0.98$, corresponding to a loss of visibility relative to Fig.~\ref{fig:spectrum}(a) of only $\approx 2$\%.

\section{Parameter considerations}\label{sec:parameters}

In this section, we discuss practical constraints on the simulation protocol proposed in this work. These constraints arise both from errors scaling with the number of register spins $N$ and from errors that accumulate with the total simulation time $T_{\mathrm{tot}}$.

\subsection{Errors due to the Floquet-Magnus expansion}
The parameters $N$, $h$, and $J$ will ultimately be constrained by the smallest achievable Floquet period $T$, which is itself controlled by how quickly single- and two-qubit gates can be performed (but which, notably, is independent of $N$). Since $E_{\mathrm{max}}=(1/2)\sum_\alpha \epsilon_\alpha$ [cf.~Eq.~\eqref{e-j}], convergence of the Floquet-Magnus expansion can be ensured by requiring that  $\sum_\alpha \epsilon_\alpha T<2\pi$. Following from the Cauchy--Schwarz inequality, the sum over $\epsilon_\alpha$ can be bounded as
\begin{align}
    \sum_\alpha \epsilon_\alpha&\leq\sqrt{N\sum_{\alpha}\epsilon_\alpha^2}\\&=\sqrt{\frac{N}{2}\mathrm{Tr}\{H_{\mathrm{BdG}}^2\}}.
\end{align}
Therefore, since $\mathrm{Tr}\{H_{\mathrm{BdG}}^2\}=8Nh^2+8J^2(N-1)$, a simple criterion for convergence of the Floquet-Magnus expansion can be written in terms of Hamiltonian parameters as
\begin{equation}\label{requirement}
    NT\sqrt{h^2+J^2\left(1-\frac{1}{N}\right)}<\pi.
\end{equation}

For fixed values of $h$, $J$, and $T$, Eq.~\eqref{requirement} leads to an upper bound on the number of register spins $N$.  Alternatively, given fixed values of $N$ and $T$, Eq.~\eqref{requirement} can be satisfied by reducing $h$ and $J$. These parameters cannot be tuned arbitrarily low, however, since the single-particle energies $\epsilon_\alpha$ [Fig.~\ref{fig:spectrum}(b)] must be spectrally resolvable on the timescale over which $\langle\sigma_x\rangle_{nT}$ is measured. Lowering $h$ and $J$ to accommodate larger values of $N$ would therefore place more stringent demands on, e.g., the CPMG coherence time of the control qubit [cf.~Eq.~\eqref{chi(nT)}]. For the parameters used in Fig.~\ref{fig:spectrum}(d), we have $E_{\mathrm{max}}T\approx 0.43\pi$, while the left-hand-side of Eq.~\eqref{requirement} gives $0.46\pi$. 

A more quantitative bound on the errors expected from the higher-order terms in the Floquet-Magnus expansion can be derived by considering [cf.~Eq.~\eqref{sigma-plus-even-odd}]
\begin{equation}\label{O-true}
    \mathcal{O}_{\mathrm{true}}(nT)=\frac{1}{2^N}\mathrm{Re}\:\mathcal{K}^n\mathrm{Tr}_{\mathrm{R}}\{\tilde{U}_\uparrow^\dagger(nT)\tilde{U}_\downarrow(nT)\},
\end{equation}
where $\tilde{U}_\uparrow(T)=e^{-iH_{\mathrm{F},\uparrow}T}\equiv U_{\mathrm{F}}$, and where, as before, $\tilde{U}_\downarrow(nT)$ has eigenstates $\ket{j}$ given by the eigenstates of $H_{\mathrm{TFIM}}$. Taking the trace in the eigenbasis of $H_{\mathrm{TFIM}}$ then gives
\begin{equation}
    \mathcal{O}_{\mathrm{true}}(nT)=\mathcal{O}_{\mathrm{ideal}}(nT)+\delta\mathcal{O}(nT),
\end{equation}
where $\mathcal{O}_{\mathrm{ideal}}(nT)=\prod_{\alpha}\cos{\left(\epsilon_\alpha nT/2\right)}$ [cf.~Eqs.~\eqref{DQC1} and \eqref{sum-to-product}], and where
\begin{align}
    &\delta \mathcal{O}(nT)=\frac{1}{2^N}\mathrm{Re}\:\mathcal{K}^n\sum_j \kappa_j^*(n) e^{-iE_j nT},\label{delta-O}\\
    &\kappa_j(n)=\langle j\vert U_{\mathrm{F}}^n\vert j\rangle-1.
\end{align}
Using the triangle inequality, the error $\delta\mathcal{O}$ on the qubit expectation value can then be bounded as 
\begin{equation}\label{bound}
    \vert \delta \mathcal{O}(nT)\vert\leq \frac{1}{2^N}\sum_j \vert \kappa_j(n)\vert.
\end{equation}

The right-hand side of Eq.~\eqref{bound} will increase with $n$ as errors due to higher-order terms in the Floquet-Magnus expansion accumulate in time. For a given set of Hamiltonian parameters, the sum $\sum_j\vert \kappa_j(n)\vert$ could be evaluated numerically by constructing and then exponentiating the one-period Floquet propagator $U_{\mathrm{F}}$. This procedure then provides a rough upper bound on the errors expected after $n$ Floquet cycles, albeit at a computational cost that grows exponentially with the system size. For the simulation results shown in Fig.~\ref{fig:spectrum}(a), for instance, the effects of higher-order terms have been included and lead, for $h=1.25J$ and $n=500$, to errors of $\approx 4\%$ relative to the ideal case (in which $H_{\mathrm{F},\uparrow}=0$ to all orders). Despite these errors, the peak positions in Figs.~\ref{fig:spectrum}(c,d) display good agreement with the expected single-particle spectrum.

In practice, the increase of $\vert\delta\mathcal{O}(nT)\vert$ with $n$ will limit the total time $T_{\mathrm{tot}}=N_{\mathrm{max}}T$ over which the dynamics can be measured before corrections due to higher-order terms  overwhelm the ideal signal. To estimate the upper bound on $T_{\mathrm{tot}}$, we perform a short-time cumulant expansion of $\kappa_j(n)$, giving 
\begin{equation}\label{O-true-2}
    \mathcal{O}_{\mathrm{true}}(nT)\simeq\frac{1}{2^N}\mathrm{Re}\:\mathcal{K}^n\sum_j e^{-i(E_j+\delta E_j)nT-\frac{1}{2}\gamma_j^2 (nT)^2},
\end{equation}
where $\delta E_j=\langle j\vert H_{\mathrm{F},\uparrow}\vert j\rangle$ represents a shift of $E_j$, and where $\gamma_j^2=\langle j\vert \Delta H_{\mathrm{F},\uparrow}^2\vert j\rangle$ depends on the variance of the Floquet Hamiltonian with respect to the TFIM eigenstate $\ket{j}$. The decay rate $\gamma_j$ is therefore set by the quantum speed limit~\cite{taddei2013quantum,fogarty2020orthogonality}, which gives the maximal rate at which $H_{\mathrm{F},\uparrow}$ can drive evolution out of state $\ket{j}$. 

Translating these quantities into shifts and broadening of the single-particle energies $\epsilon_\alpha$ is not generically possible since $H_{\mathrm{F},\uparrow}$ does not map onto a quadratic fermionic Hamiltonian; this in turn breaks the clean factorization into fermionic quasimodes [Eq.~\eqref{sum-to-product}] used to identify the peaks in $\vert S(\omega)\vert$ with the single-particle spectrum. The errors due  to having $H_{\mathrm{F},\uparrow}\neq 0$ will be small, however, provided $\delta E_j$ does not lead to appreciable dynamics on the timescale $T_{\mathrm{tot}}$ of the simulation. A simple bound on $T_{\mathrm{tot}}$ is then given by 
\begin{equation}\label{upper-bound}
    T_{\mathrm{tot}}<\frac{2\pi}{\delta E_{\mathrm{max}}}, 
\end{equation}
where $\delta E_{\mathrm{max}}=\mathrm{max}_j \delta E_j$. 
This quantity can be evaluated numerically by calculating matrix elements of the Floquet Hamiltonian in the basis of TFIM eigenstates. 

While bounded above by Eq.~\eqref{upper-bound}, $T_{\mathrm{tot}}$ must simultaneously be large enough to resolve the dynamics due to $H_{\mathrm{TFIM}}$ itself. Since the typical spacing between eigenenergies decreases with $N$, resolving individual peaks in $\vert S(\omega)\vert$ requires a larger $T_{\mathrm{tot}}$ as $N$ is increased. In particular, resolving the full bulk spectrum of the Kitaev chain would require that 
\begin{align}
    T_{\mathrm{tot}}>\frac{2\pi}{\delta\epsilon_{\mathrm{min}}},\label{resolve-bulk}\\
    \delta \epsilon_{\mathrm{min}}>\gamma_{\mathrm{max}},\label{resolve-bulk-2}
\end{align}
where $\delta\epsilon_{\mathrm{min}}$ gives the smallest spacing between single-particle energies, and where $\gamma_{\mathrm{max}}=\mathrm{max}_j \gamma_j$. For the parameters used in Fig.~\ref{fig:spectrum}(c), for instance, Eq.~\eqref{resolve-bulk} gives $T_{\mathrm{tot}}\gtrsim 140$ $\mu$s. If the goal is not to resolve the full bulk spectrum, but rather to demonstrate the presence or absence of a state near zero energy, then the requirements are significantly less stringent. In this case, the relevant comparisons can be found by replacing $\delta\epsilon_{\mathrm{min}}$ by $\Delta_{\mathrm{gap}}$ in Eqs.~\eqref{resolve-bulk}-\eqref{resolve-bulk-2}.

\subsection{Imperfect gates}
Similar considerations allow us to estimate the requirements on gate fidelity. For this purpose, we focus on the CY gates used to enforce $\bar{H}_\uparrow=0$, under the assumption that two-qubit gates have lower fidelities than single-qubit gates, and we neglect the higher-order Floquet-Magnus corrections discussed above. We adopt an error model where, instead of being rotated by angle $\theta=\pi$, the register spins are systematically over- or under-rotated by angle $\delta\theta$.  Conditioned on the control qubit being in state $\ket*{\bar{\uparrow}}$, the action of an imperfect CY gate on the $j^{\mathrm{th}}$ register spin is then to map $\tau_j^x\mapsto -\cos{\delta\theta}\:\tau_j^x-\sin{\delta\theta}\:\tau_j^z$. With the gate sequence given in Eq.~\eqref{control-pulses}, a lengthy but straightforward calculation then gives
\begin{equation}
    \bar{H}_\uparrow=\delta\theta \bar{H}_\uparrow^{(1)}+\delta\theta^2\bar{H}_\uparrow^{(2)}+O(\delta\theta^3),
\end{equation}
where
\begin{align}\label{expansion-hamiltonian}
\begin{aligned}
    &\bar{H}_\uparrow^{(1)}=(h/2)\sum_j\tau_j^z,\\
    &\bar{H}_\uparrow^{(2)}=(h/4)\sum_j \tau_j^x+(J/4)\sum_{\langle j,k\rangle}\tau_j^x\tau_k^x.
\end{aligned}
\end{align}
The ideal case of frozen evolution is recovered for $\delta\theta=0$. 

The expectation value $\mathcal{O}_{\mathrm{true}}(nT)$ then takes the same form as in Eq.~\eqref{O-true}, but in this case, $\tilde{U}_\uparrow^\dagger\neq \mathbbm{1}$ follows from an imperfect cancellation of the period-averaged Hamiltonian, $\bar{H}_\uparrow\neq 0$, rather than from higher-order terms in the Floquet-Magnus expansion. A cumulant expansion valid to quadratic order in $\delta \theta$ then recovers Eq.~\eqref{O-true-2}, but with the energy shifts and decay rates given instead by $\delta E_j=\delta \theta^2\langle j\vert \bar{H}_\uparrow^{(2)}\vert j\rangle$ and $\gamma_j^2=\delta\theta^2\langle j\vert(\bar{H}_\uparrow^{(1)})^2 \vert j\rangle$. The absence of an energy shift linear in $\delta\theta$ follows from the $\mathbbm{Z}_2$ symmetry of the TFIM Hamiltonian, which ensures that the symmetry-breaking perturbation $\bar{H}_\uparrow^{(1)}$ has no nonzero diagonal matrix elements in the basis of TFIM eigenstates.

Since for $\delta\theta\ll\pi$, the gate infidelity $\varepsilon_{\mathrm{gate}}\approx \delta\theta^2$ is simply quadratic in the rotation error,  both $\delta E_j$ and $\gamma_j^2$ are linear in the gate infidelity, with a typical scaling given by $\delta E_j\sim \varepsilon_{\mathrm{gate}}N(J+h)$ and $\gamma_j \sim (\varepsilon_{\mathrm{gate}} N)^{1/2}h$. As before, we can write down a simple criterion [cf.~Eq.~\eqref{upper-bound}] for ensuring that $T_{\mathrm{tot}}$ is short enough to avoid resolving these unintended shifts of the ideal eigenenergies $E_j$. We can then re-arrange this condition to obtain a bound on the gate infidelity that can be tolerated over a fixed time $T_{\mathrm{tot}}$. To resolve the full bulk spectrum, the gate-error-induced linewidth broadening should also be small compared to $\delta\epsilon_{\mathrm{min}}$ [cf.~Eq.~\eqref{resolve-bulk-2}]. These considerations can be combined into the following requirement on $\varepsilon_{\mathrm{gate}}$,
\begin{equation}\label{gate-infidelity}
    \varepsilon_{\mathrm{gate}}<\mathrm{min}\bigg\{\frac{2\pi}{NT_{\mathrm{tot}}(J+h)},\frac{1}{N}\left(\frac{\delta\epsilon_{\mathrm{min}}}{h}\right)^2\bigg\}.
\end{equation}
For the parameters used in Fig.~\ref{fig:spectrum}, where $T_{\mathrm{tot}}=250\:\mu\mathrm{s}$, Eq.~\eqref{gate-infidelity} gives $\mathcal{\varepsilon}_{\mathrm{gate}}<0.6\%$ for $h=1.25 J$, limited by the condition involving $T_{\mathrm{tot}}$. This estimate should not be interpreted as a hard cutoff, especially since the gate-error model assumed here may not apply in a given experimental setting. The inequality in Eq.~\eqref{gate-infidelity} was also derived using a very conservative parametric dependence on $h$, $J$, and $N$, without accounting for the exact polarization of the spins along the $z$ and $x$ axes [cf.~Eq.~\eqref{expansion-hamiltonian}] in the different $\ket{j}$ eigenstates.

\subsection{Inhomogeneous broadening}
Finally, we remark that resolving individual eigenenergies requires sufficient stability of the Hamiltonian parameters themselves. In the presence of shot-to-shot variation in $h$ and $J$, the peaks in $\vert S(\omega)\vert$ will be inhomogeneously broadened. To leading order and for independent fluctuations $\delta h$ and $\delta J$, the resulting width $\sigma$ of the peak at $\omega=\epsilon_\alpha$ is given by $\sigma^2\simeq (\partial_h\epsilon_\alpha)^2\sigma_h^2+(\partial_J\epsilon_\alpha)^2\sigma_J^2$, where $\sigma_h^2$ and $\sigma_J^2$ are the variances of $\delta h$ and $\delta J$. To distinguish individual spectral peaks, the linewidth broadening should then be small relative to the minimum peak separation, $\sigma<\delta\epsilon_{\mathrm{min}}$. 

The  spacing between single-particle energies can be found from the dispersion $\epsilon_\alpha=2(h^2+J^2+2hJ\cos{k_\alpha})^{1/2}$, where, for a finite open chain, the allowed momenta $k_\alpha$ are determined by a boundary-dependent quantization condition and are generally not equally spaced~\cite{leumer2020exact}. However, away from the band edge and in the large-$N$ limit, their characteristic spacing is $\Delta k=O(1/N)$, leading to a characteristic bulk energy spacing that is also $O(1/N)$. This highlights the need for greater system stability as the number of spins is increased. As before, simply distinguishing states near zero energy from bulk states comes with relaxed requirements on peak resolution and could therefore be achieved with higher overall levels of inhomogeneous broadening.

\section{Conclusion}

In this work, we have presented a hybrid digital-analog protocol for simulating the transverse-field Ising model (TFIM) and its equivalent Kitaev chain using a $2\times N$ quantum-dot array. We showed that by encoding a control qubit in a repetition code and applying a tailored sequence of parallelized CY gates, the stroboscopic system evolution can be mapped onto the output of the one-clean-qubit (DQC1) model of computation. This Floquet-engineered approach therefore imprints complete spectral information onto the coherence of the control qubit, allowing all $2^N$ eigenenergies of the TFIM to be extracted via measurements of a single spin. The $\mathbbm{Z}_2$ symmetry of the TFIM, generated by global spin flips $\hat{P}=\prod_j (-i \tau_j^x)$, ensures that the analog portion of the simulation remains compatible with dynamical decoupling of the register spins: Since the TFIM Hamiltonian commutes with $\hat{P}$, dynamical-decoupling pulses can be used to suppress spin dephasing without averaging out the native terms driving the quantum simulation.

We also showed that classical postprocessing of the measured time-domain qubit coherence can be used to isolate the $N$ single-particle energies of the Kitaev chain. By varying the strength of the synthetic transverse field, measurements of the single-particle spectrum can then be used to map the crossover from the trivial regime to the topological regime of the Kitaev chain. Numerical simulations for $N=6$ suggest that these single-particle features will remain clearly resolvable even in the presence of realistic $1/f$ dephasing noise, due in large part to the compatibility of the simulated Hamiltonian with dynamical decoupling. By bypassing the need for simultaneous full-register readout, this approach offers a viable near-term pathway for simulating and probing the spectral characteristics of a topological crossover using present-day spin-qubit devices.

\section{Code availability}
\noindent Code supporting the simulations will be made available at a later date.

\begin{acknowledgements}
    Calculations were performed at the sciCORE scientific computing center at the University of Basel. This work was supported by the Swiss National Science Foundation, NCCR SPIN (Grant No.~225153). DL acknowledges the Deanship of Research and the Quantum Center at KFUPM for the support received under Grant no.~CUP25102 and no.~INQC2600, respectively. 
\end{acknowledgements}

\appendix

\section{Realization of the TFIM in a rotating frame}\label{appendix:rot-frame}

The result for $\langle\sigma_+\rangle_{nT}$ given in Eq.~\eqref{DQC1} was obtained under the assumption that the Floquet Hamiltonian $H_{\mathrm{F},\uparrow}$ is well approximated by the period-averaged Hamiltonian $\bar{H}_\uparrow$ [Eq.~\eqref{magnus-expansion}]. Since this approximation itself requires that  $E_{\mathrm{max}} T<\pi$, engineering $H_{\mathrm{TFIM}}$ using static fields is not a viable approach due to the size of the magnetic fields typically used in experiments, which lead to qubit splittings of a few GHz. This would in turn require a sub-nanosecond Floquet period. 

The TFIM Hamiltonian can be realized in a rotating frame by subjecting all register spins to continuous Rabi driving. We take the lab-frame Hamiltonian $H_{\mathrm{lab}}$ describing the driven register spins to be given by
\begin{equation}\label{lab-hamiltonian}
    H_{\mathrm{lab}}=\sum_{i=1}^N \left(\frac{\Delta_i}{2}\tau_i^{z}+\Omega_i(t)\tau_i^{x}\right)+\sum_{i=1}^{N-1}\bm{\tau}_{i}\cdot\bm{J}\cdot\bm{\tau}_{i+1},
\end{equation}
where $\Delta_i$ is the splitting of the $i^{\mathrm{th}}$ spin and $\Omega_i(t)=\Omega_i\cos(\omega_it+\phi_i)$ is a Rabi drive at frequency $\omega_i$. The second term in Eq.~\eqref{lab-hamiltonian} models an exchange interaction between neighboring spins, which we allow to be anisotropic as is commonly the case for hole spins~\cite{geyer2024anisotropic}.  Heisenberg exchange is recovered when $\bm{J}\propto\mathrm{diag}(1,1,1)$.  Although the strength of the exchange is taken to be the same for all pairs of neighboring spins, the generalization to unequal exchange is straightforward.

We define a rotating frame obtained through the unitary transformation  $U_{\mathrm{drive}}(t)=\prod_j e^{-i (\omega_j/2)\tau_j^z t}$. In this frame, $H_{\mathrm{lab}}$ is given, up to a rotating-wave approximation, by
\begin{align}
    &H_{\mathrm{RWA}}= \sum_{i=1}^N H_i+J_{zz}\sum_{i=1}^{N-1}\tau_i^{z}\tau_{i+1}^{z},\label{rot-frame-1}\\
    &H_i=\frac{\delta_i}{2}\tau_i^{z}+\frac{\Omega_i}{2}(\cos{\phi_i}\:\tau_i^{x}+\sin{\phi_i}\:\tau_i^{y}),\label{rot-frame-2}
\end{align}
where $\delta_i=\Delta_i-\omega_i$ is the detuning between the splitting of the $i^{\mathrm{th}}$ spin and the frequency of its Rabi drive. The requirements for validity of the rotating-wave approximation made above are that 
\begin{align}
    &\Omega_i\ll\omega_i,\quad \forall i,\\
    &J_{\alpha\beta}\ll \vert \omega_i-\omega_{i+1}\vert,\quad(\alpha,\beta)\neq (z,z).\label{rwa-2}
\end{align}
The first of these conditions suppresses counter-rotating terms oscillating at $\pm2\omega_i$, while the second suppresses all nonsecular terms in the exchange interaction. Setting $\delta_i=0$, $\Omega_i=h$, $\phi_i=0$, and $J_{zz}=J$, we then recover the TFIM Hamiltonian given in Eq.~\eqref{H0}. Since obtaining rotating-frame Ising interactions using this approach may rely on having different spin splittings [cf.~Eq.~\eqref{rwa-2}], the use of a  global driving field~\cite{seedhouse2021quantum, hansen2021pulse,vahapoglu2022coherent} is best suited to the case where the lab-frame exchange is natively Ising and all spins are degenerate (in which case $\delta_i=0$ can be satisfied simultaneously); small inhomogeneities could be tolerated provided they do not impact the system dynamics on the timescale $T_{\mathrm{tot}}$ of the simulation. In other cases, resonant Rabi driving could instead be applied to individual spins using electric-dipole spin resonance~\cite{golovach2006electric,nowack2007coherent,pioro2008electrically}, delivered via the plunger gates used to define individual quantum dots.

Working in the rotating frame defined by $U_{\mathrm{drive}}$ has no impact on the expectation value $\langle\sigma_+\rangle_t$ of the control qubit  [Eq.~\eqref{coherence}]. This follows from the fact that $U_{\mathrm{lab}}(t)=U_{\mathrm{drive}}(t)U(t)$ with $[U_{\mathrm{drive}},\sigma_+]=0$, giving
\begin{align}
    \langle\sigma_+\rangle_t=\mathrm{Tr}\{\sigma_+\rho_{\mathrm{lab}}(t)\}=\mathrm{Tr}\{\sigma_+\rho(t)\},
\end{align}
where $\rho_{\mathrm{lab}}(t)=U_{\mathrm{lab}}(t)\rho(0)U_{\mathrm{lab}}^\dagger(t)$ is the lab-frame density matrix, and where the rotating-frame density matrix $\rho(t)=U(t)\rho(0)U^\dagger(t)$ evolves under the Hamiltonian $H(t)$ given in Eq.~\eqref{rot-frame-hamiltonian}, with $H_{\mathrm{TFIM}}$ engineered from $H_{\mathrm{RWA}}$ as just described.

Two-qubit gates [described by the term $\ketbra*{\bar{\uparrow}}\otimes\hat{H}_{\mathrm{c}}(t)$ in Eq.~\eqref{hamiltonian-pi-frame}] can be realized by turning on the exchange interaction between spins in the top and bottom rows of the $2\times N$ array. We write the lab-frame Hamiltonian describing the exchange interaction between the $i^{\mathrm{th}}$ control spin (having splitting $\omega_{\mathrm{q},i}$) and the $j^{\mathrm{th}}$ register spin as
\begin{equation}\label{exchange-interrow}
    H_{\mathrm{ex}}=\bm{\sigma}_i\cdot \bm{J}\cdot\bm{\tau}_j,
\end{equation}
where we again denote the exchange tensor by $\bm{J}$, with the implicit understanding that the exchange interaction in Eq.~\eqref{exchange-interrow} can be tuned independently from the exchange between pairs of register spins [cf.~Eq.~\eqref{lab-hamiltonian}].

\begin{figure}
    \centering
    \includegraphics[width=0.8\linewidth]{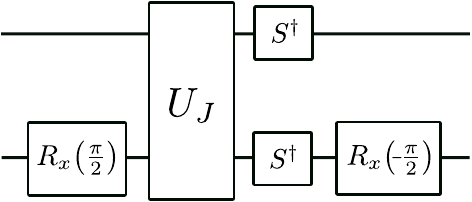}
    \caption{Rotating-frame CY gate compiled with hardware-native operations. The entangling gate $U_J$ can be realized by  turning on the exchange interaction between the target register spin and any spin involved in the repetition code used to define the control qubit. The rotations  $R_x(\varphi)$ and $S^\dagger$ of the register spin could be achieved by pulsing the amplitude and chirping the phase of the Rabi drive applied to that spin. Since the CY gate preserves the state of the control qubit, the $S^\dagger$ gate on the control qubit can simply be tracked in software.}
    \label{fig:native-gates}
\end{figure}

Notably, there are terms in $H_{\mathrm{ex}}$ that do not preserve the logical subspace $\mathrm{span}\{\ket*{\bar{\uparrow}},\ket*{\bar{\downarrow}}\}$ of the  control qubit. To transform to the frame in which Eq.~\eqref{hamiltonian-pi-frame} is written, we consequently take $U_\mathrm{0}(t)=U_{\mathrm{DD}}(t)U_\mathrm{q}(t)$ [Eq.~\eqref{qubit-transformation}] to act on the full Hilbert space of the spins encoding the control qubit. This is achieved by taking 
\begin{align}
    &U_\mathrm{q}(t)=\prod_{j=1}^{\lceil N/2\rceil}e^{-\frac{i}{2}\phi_j(t)\sigma_j^z},\\
    &U_{\mathrm{DD}}(t)=\big[\prod_{j=1}^{\lceil N/2\rceil} (-i\sigma_j^x)\big]^{m(t)}\hat{P}^{\ell(t)},\label{U-dd-full}
\end{align}
where $\phi_j(t)=\omega_{\mathrm{q},j}\int_0^tdt's(t')$ and $s(t)=(-1)^{m(t)}$. The relation $\sigma_x=\prod_j \sigma_j^x$ [cf.~Eq.~\eqref{pi-pulses}] used to write Eq.~\eqref{U-dd-full} follows from the fact that the logical X gate in a repetition code is given by a product of X gates acting on all spins in the code. When restricted to the logical subspace, the unitary $U_{\mathrm{q}}(t)$ can be written in terms of the control-qubit operator $\sigma_z$ (as was done in the main text) since the splitting $\omega_{\mathrm{q}}=\sum_j \omega_{\mathrm{q},j}$ of the repetition-code qubit is simply the sum of the physical spin splittings.

We transform $H_{\mathrm{ex}}$ to the frame obtained via the unitary transformation $U_{\mathrm{drive}}(t)U_0(t)$, allowing for a direct comparison with the rotating-frame Hamiltonian given in Eq.~\eqref{hamiltonian-pi-frame}. In this frame,  $\sigma_i^\pm$ and $\tau_j^\pm$ oscillate at frequencies $\pm\partial_t \phi_i=\pm \omega_{\mathrm{q},i} s(t)$ and $\pm\omega_j$, respectively. Hence, provided $J\ll\omega_{\mathrm{q},i},\omega_j,\vert \omega_{\mathrm{q},i}-\omega_j\vert$, performing a rotating-wave approximation on the exchange interaction gives
\begin{equation}\label{h-ex-RWA}
    \hat{H}_{\mathrm{ex,RWA}}(t)= s_{\Sigma}(t)J_{zz}\sigma_i^z\tau_j^z,
\end{equation}
where the sign function $s_{\Sigma}(t)=(-1)^{m(t)+\ell(t)}$ arises from the dynamical-decoupling pulses applied to the control qubit and register spin. Within this rotating-wave approximation, the exchange Hamiltonian [Eq.~\eqref{h-ex-RWA}] preserves the logical subspace spanned by $\ket*{\bar{\uparrow}},\ket*{\bar{\downarrow}}$. Its action on the control qubit can therefore be obtained by projecting $\hat{H}_{\mathrm{ex,RWA}}$ into the logical subspace via the projection operator $\hat{P}_{\mathrm{Q}}=\sum_{\alpha=\uparrow,\downarrow}\ketbra*{\bar{\alpha}}$:
\begin{equation}\label{logical-CZ}
    \hat{P}_{\mathrm{Q}}\hat{H}_{\mathrm{ex,RWA}}(t)\hat{P}_{\mathrm{Q}}= s_{\Sigma}(t)J_{zz}\sigma_z\tau_j^z.
\end{equation}

This Hamiltonian can be used to perform gates on the control qubit and $j^{\mathrm{th}}$ register spin:
By turning on the exchange for a time $\pi/(4J_{zz})$, the two-qubit basis states will acquire phases of $\pm \pi/4$ conditioned on their $ZZ$ parity. For $s_\Sigma(t)=+1$, this yields an entangling gate $U_J$, which, in the basis $\{\ket*{\bar{\downarrow}\downarrow}, \ket*{\bar{\downarrow}\uparrow}, \ket*{\bar{\uparrow}\downarrow}, \ket*{\bar{\uparrow}\uparrow}\}$, takes the form
\begin{equation}
    U_J=\begin{pmatrix}
        1 & 0 & 0 & 0\\
        0 &  i & 0 & 0\\
        0 & 0 &  i & 0\\
        0 & 0 & 0 & 1
    \end{pmatrix},
\end{equation}
up to a global phase.
This gate is equivalent to a CZ gate up to single-spin rotations: 
\begin{align}\label{cz-equivalence}
    &\mathrm{CZ}=(S^\dagger\otimes S^\dagger)U_J,\\
    &S=\begin{pmatrix}
        1&0\\
        0&i
    \end{pmatrix}.
\end{align}
Due to the alternation in the sign of the exchange coupling in Eq.~\eqref{logical-CZ}, realizing the same CZ gate for $s_\Sigma(t)=-1$ requires that the correction $(S\otimes S)$ be applied in place of $(S^\dagger\otimes S^\dagger)$ [cf.~Eq.~\eqref{cz-equivalence}].

Given CZ, the required CY gate can be realized by rotating the register spin about the $x$ axis by angles $\pi/2$ and $-\pi/2$ before and after the CZ gate, respectively (Fig.~\ref{fig:native-gates}). Single-spin rotations of the $j^{\mathrm{th}}$ register spin can be realized by pulsing the strength of its Rabi drive: If the drive amplitude $\Omega_j$ is temporarily increased so that it exceeds the strength of all other terms in $H_{\mathrm{RWA}}$ [Eq.~\eqref{rot-frame-1}], then the Rabi drive produces a strong control pulse that rotates the spin about the $x$ axis in the rotating frame (for $\phi_i=0$). Any dynamics due to exchange interactions can be neglected during this rotation provided the gate time is short relative to the timescale for free evolution under these interactions. Similarly, phase gates like $S$ can be implemented by chirping the frequency $\omega_j$ of the Rabi drive: By temporarily increasing the drive detuning $\delta_j$ of the $j^{\mathrm{th}}$ spin so that it dominates over all other terms in $H_{\mathrm{RWA}}$, rotations about the $z$ axis can be realized as well.

\end{document}